\documentclass[%
reprint,
superscriptaddress,
amsmath,amssymb,
aps,
prd,
longbibliography,
]{revtex4-2}

\usepackage{graphicx,wrapfig,lipsum}
\usepackage{subeqnarray}
\usepackage{url,hyperref}
\usepackage{color}
\usepackage[normalem]{ulem}
\usepackage{verbatim}
\usepackage{bm}

\usepackage{natbib}
\newcommand{\req}[1]{Eq.\,({\ref{#1}})}

\usepackage{xcolor}
\colorlet{editcyan}{cyan!80!black} 
\colorlet{EDITCYAN}{editcyan} 

\begin{document}
\title{Attractor Trajectories in Particle Dynamics Subject to Radiation Reaction}
\author{Martin S. Formanek}
\email{martin.formanek@eli-laser.eu}
\affiliation{ELI Beamlines Facility, The Extreme Light Infrastructure ERIC, 252 41 Dolní Břežany, Czech Republic}
\author{Johann Rafelski}
\email{johannr@arizona.edu}
\affiliation{Department of Physics, The University of Arizona, 85721 Tucson, AZ, USA}

\date{November 30, 2026}

\begin{abstract}
Charged-particle motion in an external electromagnetic field is analyzed for the analytically soluble example of Landau--Lifshitz radiation-reaction friction in quasi-constant and quasi-homogeneous fields. We show that in the instantaneous particle rest frame the field energy density $\mathcal{E}_u$ approaches $\mathcal{M}c^2$, set by the Lorentz-invariant field mass density $\mathcal{M}$. Equivalently, the energy flux of the external field observed by the particle monotonically approaches zero. Using these insights, we characterize the dynamical evolution of all possible particle trajectories and show how motion evolves to be radiation friction-free: in a purely magnetic field the particle approaches inertial motion, and in any other non-null field it approaches motion along a Rindler trajectory. In a null field the proper acceleration decays to zero and the motion asymptotically approaches the null direction of the field.
\end{abstract}
\maketitle
\section{Introduction}\label{sec:Intro}
An accelerated particle with mass $m$ and charge $q$ radiates, and the radiation reaction (RR) force of this vacuum friction acts back on particle motion by recoil against the emitted radiation. Exploration of forces that create relevant strong acceleration is a domain of physics of intense interest, allowing studies of physical laws at the acceleration frontier~\cite{Steinmetz2027}. In classical electrodynamics RR is described by the Lorentz--Abraham--Dirac (LAD) equation~\cite{Dirac:1938nz}. In practice, one usually works with its reduced-order Landau--Lifshitz (LL) form~\cite{Landau:1975pou}, which agrees with both LAD and the alternate Eliezer--Ford--O'Connell (EFO) formulation~\cite{Eliezer:1948qkl,Ford:1991ntp} to first order in external Lorentz force~\cite{Price:2021zqq}. Recently, interest in RR has been renewed by laser facilities whose fields are strong enough to probe the quantum regime of the interaction~\cite{DiPiazza:2011tq,Blackburn:2019rfv,Gonoskov:2021hwf,Fedotov:2022ely} as the first experiments show~\cite{Poder:2017dpw,Cole:2017zca,Mirzaie:2024iey,Los:2024ysw}. Beyond fundamental questions proper understanding of radiation emission in laser–matter interactions is essential for successfully implementing laser-driven inertial confinement fusion \cite{lindl1995development,craxton2015direct}.

A known feature of vacuum friction is that it vanishes to all orders on particular trajectories: i) for inertial motion in vacuum, and ii) for the case of a Rindler motion along a straight line with constant proper acceleration~\cite{Born:1909rpp,Rindler:1960zz}. Along such a trajectory, the charge radiates at the Larmor rate, yet the LAD force vanishes identically~\cite{FultonRohrlich1960,Boulware:1979qj,Rohrlich2007,EriksenGron2000}, while the radiated energy is supplied by the Schott term~\cite{Teitelboim:1970mw}. A charge starting from rest in a constant electric field moves on exactly such a worldline. In a general constant field, however, the Lorentz-force trajectories combine Rindler hyperbolic motion with a cyclotron-like gyration and RR does
not vanish~\cite{Taub1948,Dumais1985,Fradkin:1978kx,Munoz1997,Chin2009}. 

In this letter, we present a new path for understanding the vacuum friction and resolve the question of natural dynamical evolution to RR-free trajectories in the process. Our approach relies on the fact that the energy flux of the external field seen by the moving charge can be expressed in covariant form within Maxwell theory of electromagnetism using two building blocks: i) the energy density of the field in the rest frame of the charge $\mathcal{E}_u$ and ii) the Lorentz-invariant mass density of the field $\mathcal{M}$ constructed from field invariants $S$ and $P$. For the example of quasi-constant quasi-homogeneous external fields and vacuum friction described by the LL reaction force, we show that $\mathcal{E}_u$ obeys a simple analytical equation 
\begin{equation}
\label{eq:introEvol}
\dot{\mathcal{E}}_u = -\frac{2\tau_0 q^2}{\varepsilon_0 m^2 c^2}
\left(\mathcal{E}_u^2-\mathcal{M}^2c^4\right),
\end{equation}
where $\tau_0=q^2/(6\pi\varepsilon_0 mc^3)$ is the RR time scale, $\varepsilon_0$ is the vacuum permittivity, $c$ is the speed of light, and over-dot denotes the proper time derivative. 

We demonstrate that the right-hand side of \req{eq:introEvol} is proportional to the squared Poynting flux of the external field seen by the charge. This implies that the vacuum friction, as described by the LL reaction force, drives the charge towards a state in which it sees no flux of the external field. Our review of the prior work~\cite{Bulanov2011,Gonoskov:2017lyz,Samsonov:2018skj,Ekman:2021vwg,Cai:2022mkw,Cai:2023ykg,Samsonov2023} suggests that this is a general result demonstrated for the first time here. This insight is further supported by demonstration that the flux-free four-velocities are the eigenvectors of the energy-momentum tensor $T^{\mu}_{\nu}$, for which the vacuum friction vanishes identically. As a consequence, for charged-particle motion in any constant non-null field $(S^2 + P^2 \neq 0)$ the trajectory approaches the Rindler motion, unless the field is purely magnetic, in which case it approaches the axial inertial force-free motion. Considering the special case of a crossed null field ($S^2+P^2=0$) which is the constant-field limit of a plane wave and approximates ultrarelativistic behavior in any field, particle motion approaches a uniform lightlike motion.

To classify the motion of charged particles in constant electromagnetic fields subjected to RR in LL form, the eigenspace projectors of the electromagnetic field tensor squared $(F^2)^\mu_\nu$ are used~\cite{Yaremko:2013eqe}. Previous works have studied the late-time asymptotics of the physical solutions of the LAD and LL equations in constant fields~\cite{Kazinski:2010ce}, and in plane wave fields~\cite{Kazinski:2013vga}. The tendency of the charge to locally move towards a direction with no lateral acceleration (so-called radiation-free direction) has also been described, see~\cite{Bulanov2011,Gonoskov:2017lyz,Samsonov:2018skj,Ekman:2021vwg,Samsonov2023}. Here, we opt to use the term RR-free (spelled out: radiation-reaction-force-free) direction, as the accelerated particle can still radiate through the Larmor term by changing the magnitude of velocity, but with no effect on the shape of its trajectory. 

Manuscript organization: In Sec.~\ref{sec:flux} we derive \req{eq:introEvol} and its flux form. In Sec.~\ref{sec:nonnull} we treat non-null fields with $S^2+P^2\neq0$ using the eigenspace decomposition of $(F^2)^\mu_\nu$ which gives the proper time evolution of the four-velocity explicitly and shows the approach to the Rindler or force-free motion. Null crossed fields ($S^2+P^2=0$) are addressed in Sec.~\ref{sec:cross}, both exact and as an approximation of ultrarelativistic motion. In this case the proper acceleration decays to zero and the velocity tends to $c$ along the null direction. In Sec.~\ref{sec:examples} we discuss the approach to the attractor trajectories in constant fields of Sec.~\ref{sec:nonnull} and evaluate the decay of perpendicular momentum with reference to RR-free trajectories. We close with a discussion of future extensions to non quasi-constant fields and alternate models of RR in Sec.~\ref{sec:discussion}.

Notation: We work in flat Minkowski metric $\eta_{\mu\nu} = \text{diag}(1,-1,-1,-1)$ using SI units throughout. The worldline $z^\mu(\tau)$ is parametrized by particle proper time $\tau$. It follows that $u^\mu=\dot z^\mu$ is the particle four-velocity with $u^2=c^2$, and the particle four-acceleration $a^\mu=\dot u^\mu$ satisfies $u\cdot a=0$ and $a^2\leq 0$.

\section{Evolution of the field flux seen by the charge}\label{sec:flux}
The energy-momentum tensor of the external field $T^\mu_\nu$ contracted with the four-velocity has, in the rest frame of the particle $u^\mu = (c, 0)$, components
\begin{equation}\label{eq:tmnurf}
 T^\mu_\nu u^\nu|_\text{RF} = c(T^{00}, T^{i0}) = (c \mathcal{E}_u, \boldsymbol{\Phi}_u)\,,
\end{equation}
where $\boldsymbol{\Phi}_u$ is the energy flux of the field as seen by the particle and 
\begin{equation}\label{eq:Eu}
 \mathcal{E}_u \equiv \frac{u \cdot T \cdot u}{c^2}\,,
\end{equation}
is the covariant definition of the field energy density measured by the particle in its rest frame. Squaring Eq.~(\ref{eq:tmnurf}) gives a Lorentz-invariant left-hand side, so
\begin{equation}
 u \cdot T^2 \cdot u= c^2 \mathcal{E}_u^2-\Phi_u^2\,,
\end{equation}
where $\Phi_u \equiv |\boldsymbol{\Phi}_u|$ is the field flux magnitude. In theories where $(T^2)^\mu_\nu \propto \delta^\mu_\nu$, which is the case of Maxwellian Electrodynamics, the left-hand side can be used to define the invariant field mass density~\cite{Rafelski:2017hyt}
\begin{equation}\label{eq:E}
 \mathcal{M} \equiv \frac{1}{c^3}\sqrt{u \cdot T^2 \cdot u}\,,
\end{equation}
independent of the particle four-velocity $u^\mu$. Then the magnitude of the field flux seen by the particle is
\begin{equation}\label{eq:flux}
 \Phi_u \equiv c\sqrt{\mathcal{E}_u^2-\mathcal{M}^2c^4}\,.
\end{equation}
This means that $\mathcal{E}_u \geq \mathcal{M}c^2$ with the equality holding for a particle which sees no field flux. The change of the energy flux in a constant field reads
\begin{equation}\label{eq:dflux}
 \dot{\Phi}_u = \frac{c^2}{\Phi_u}\mathcal{E}_u \dot{\mathcal{E}}_u = \frac{2}{\Phi_u}\mathcal{E}_u (u\cdot T \cdot a)\,.
\end{equation}

Specifically, for the energy-momentum tensor of the external Maxwell electromagnetic field
\begin{equation}\label{eq:Tmunu}
 T^\mu_\nu \equiv c^2\varepsilon_0\left( F^{\mu\alpha}F_{\alpha\nu} + S \delta^\mu_\nu\right)\,,
\end{equation}
the square is proportional to the Kronecker delta
\begin{equation}
 (T^2)^\mu_\nu = (c^2 \varepsilon_0)^2(S^2 + P^2)\delta^\mu_\nu\,,
\end{equation}
since $(F^4)^\mu_\nu = -2S (F^2)^\mu_\nu + P^2 \delta^\mu_\nu$ and where the field invariants are defined as
\begin{align}
 S &\equiv \frac{1}{4} F^{\mu\nu}F_{\mu\nu} = \frac{1}{2}(B^2 - E^2/c^2)\,,\\
 P &\equiv \frac{1}{4} F^{\mu\nu} \widetilde{F}_{\mu\nu} = \boldsymbol{E} \cdot \boldsymbol{B}/c\,,
\end{align}
with $\widetilde{F}_{\mu\nu} = \frac{1}{2}\epsilon_{\mu\nu\alpha\beta}F^{\alpha\beta}$ being the dual EM tensor, $\epsilon_{0123}=+1$. The invariant mass density of the field is then given by Eq.~(\ref{eq:E})
\begin{equation}\label{eq:invenergy}
 \mathcal{M} = \varepsilon_0 \sqrt{S^2+P^2}\,.
\end{equation} 
The four-acceleration in Eq.~(\ref{eq:dflux}) can be replaced by the particle dynamics. As the simplest nontrivial example, the four-acceleration given by the LL equation of motion reads
\begin{equation}\label{eq:LL}
 a^\mu = \frac{q}{m}F^\mu_\nu u^\nu + \tau_0 \frac{q}{m}\dot{F}^\mu_\nu u^\nu +\tau_0 \frac{q^2}{m^2} P^\mu_\nu (F^2)^\nu_\alpha u^\alpha\,,
\end{equation}
where $P^\mu_\nu \equiv \delta^\mu_\nu - u^\mu u_\nu/c^2$ is the orthogonal projector for four-velocity and the second term is dropped in constant fields. When substituted into $\dot{\mathcal{E}}_u = 2 u \cdot T \cdot a/c^2$ the Lorentz force term vanishes due to antisymmetry of $F^\mu_\nu$; a particle in constant EM fields in the absence of RR sees constant field energy density in its rest frame. The rest is
\begin{equation}
\begin{split}
 \dot{\mathcal{E}}_u &= \frac{2\tau_0 q^2}{m^2 c^2}\left(u\cdot T\cdot F^2 \cdot u - \frac{u\cdot T \cdot u}{c^2}u \cdot F^2\cdot u\right)\\
 &= \frac{2\tau_0 q^2}{\varepsilon_0 m^2 c^2}\left(\frac{u\cdot T^2 \cdot u}{c^2}- \frac{(u \cdot T \cdot u)^2}{c^4} \right)\,,
\end{split}
\end{equation}
where we substituted $(F^2)^\mu_\nu$ from Eq.~(\ref{eq:Tmunu}). Finally, identifying $\mathcal{E}_u$ and $\mathcal{M}$
\begin{equation}\label{eq:energydensityrelax}
 \dot{\mathcal{E}}_u = - \frac{2\tau_0 q^2}{\varepsilon_0m^2 c^2}\left(\mathcal{E}_u^2 - \mathcal{M}^2 c^4\right)\,.
\end{equation}
The LL RR force acts to relax the energy density of the field as seen by the moving particle $\mathcal{E}_u$ so that it asymptotically tends to the invariant value $\mathcal{M}c^2$. When used in Eq.~(\ref{eq:dflux})
\begin{equation}\label{eq:fluxValue}
 \dot{\Phi}_u = - \frac{2\tau_0 q^2}{\varepsilon_0 m^2 c^2} \mathcal{E}_u\Phi_u\,.
\end{equation}
The instantaneous rate of change of the natural logarithm of the field flux seen by the particle is given by the observed field energy density $\mathcal{E}_u$.

\section{Motion in fields with \texorpdfstring{$\bm{S^2+P^2\neq 0}$}{}}\label{sec:nonnull}
\subsection{Decomposition of motion into eigenspaces}
The eigenvalues of the tensor $(F^2)^\mu_\nu$ are~\cite{Dumais1985,Munoz1997}
\begin{equation}\label{eq:eigenvalues}
 \Omega_{\mathcal{A},\mathcal{B}} = -S \pm \sqrt{S^2 + P^2}\,,
\end{equation}
each doubly degenerate. It is advantageous to introduce two new variables $\mathcal{A}$ and $\mathcal{B}$ so that
\begin{align}
 \mathcal{A}^2 &\equiv -S + \sqrt{S^2+P^2}\,,\\
 \mathcal{B}^2 &\equiv S + \sqrt{S^2+P^2}\,.
\end{align}
The inverse relationship is given by 
\begin{equation}
 S = \frac{1}{2}(\mathcal{B}^2-\mathcal{A}^2)\,, \quad |P| = \mathcal{A}\mathcal{B}\,.
\end{equation}
The variables $\mathcal{A}$ and $\mathcal{B}$ are the invariant field strengths familiar from the Euler--Heisenberg--Schwinger effective action~\cite{Schwinger:1951nm}.
For a purely electric field $\mathcal{A} = |\boldsymbol{E}|/c$ and $\mathcal{B} = 0$. On the contrary, for a pure magnetic field $\mathcal{A} = 0$ and $\mathcal{B} = |\boldsymbol{B}|$. In terms of these variables, the eigenvalues given in Eq.~(\ref{eq:eigenvalues}) are simply
\begin{equation}
 \Omega_\mathcal{A} = \mathcal{A}^2\,, \quad \Omega_\mathcal{B} = - \mathcal{B}^2\,.
\end{equation}

The projectors on the corresponding eigenspaces are~\cite{Fradkin:1978kx}
\begin{align}
 (\Pi_\mathcal{A})^\mu_\nu &= \frac{F^{\mu\alpha}F_{\alpha\nu} + \mathcal{B}^2 \delta^\mu_\nu}{\mathcal{A}^2 + \mathcal{B}^2}\,,\label{eq:projA}\\
 (\Pi_\mathcal{B})^\mu_\nu &= \frac{-F^{\mu\alpha}F_{\alpha\nu} + \mathcal{A}^2 \delta^\mu_\nu}{\mathcal{A}^2 + \mathcal{B}^2}\,.\label{eq:projB}
\end{align}
Such projectors exist for every field in which the denominator is nonzero $\mathcal{A}^2+\mathcal{B}^2 = 2\sqrt{S^2 + P^2}\neq0$. Both traces are
\begin{equation}
 (\Pi_\mathcal{A})^\mu_\mu = (\Pi_\mathcal{B})^\mu_\mu = 2\,.
\end{equation}
Therefore both subspaces are two-dimensional. We will use subscripts $\mathcal{A}$ or $\mathcal{B}$ for the projected four-vector quantities. For example, we can decompose the four-velocity as
\begin{equation}
 u^\mu = u^\mu_\mathcal{A} + u^\mu_\mathcal{B}\,,
\end{equation}
\begin{equation}
 u_\mathcal{A}^\mu \equiv (\Pi_\mathcal{A})^\mu_\nu u^\nu\,, \quad u_\mathcal{B}^\mu \equiv (\Pi_\mathcal{B})^\mu_\nu u^\nu\,.
\end{equation}

We introduce the variable $\kappa \geq 0$ that measures how much the four-velocity of the charge is outside the $\mathcal{A}$ plane. This can be done by introducing the magnitudes
\begin{align}
u^2_\mathcal{A} &=c^2+\kappa\,,\label{eq:uA2}\\
u^2_\mathcal{B} &= - \kappa\,,\label{eq:uB2}
\end{align}
so that $u^2 = u^2_\mathcal{A} + u^2_\mathcal{B} = c^2$ is preserved. The proper time derivative of $\kappa$ reads
\begin{equation}
 \dot{\kappa} = - \frac{d}{d\tau} u_\mathcal{B}^2 = - 2 u_\mathcal{B} \cdot a_\mathcal{B} = - 2 u_\mathcal{B} \cdot a\label{eq:dotr}\,.
\end{equation}
As $\kappa \rightarrow 0$ the four-velocity becomes confined to the $\mathcal{A}$ subspace. 

\subsection{LL radiation reaction effect}
The field energy density as seen by the charge is
\begin{equation}\label{eq:EuSPnonzero}
 \mathcal{E}_u(S^2 + P^2 \neq 0) = \mathcal{M}c^2\left(1 + \frac{2\kappa}{c^2}\right)\,,
\end{equation}
where the invariant mass density of the field Eq.~(\ref{eq:invenergy}) is
\begin{equation}\label{eq:ESPnonzero}
\begin{split}
 \mathcal{M}(S^2+P^2\neq 0) &= \varepsilon_0 \sqrt{S^2 + P^2}\\
 &= \frac{1}{2}\varepsilon_0(\mathcal{A}^2 + \mathcal{B}^2)\,.
\end{split}
\end{equation}
Substituting both into Eq.~(\ref{eq:energydensityrelax}) yields
\begin{equation}\label{eq:dotkappa}
 \dot{\kappa }= - \frac{4\tau_0q^2}{\varepsilon_0 m^2}\mathcal{M}\kappa \left(1+\frac{\kappa}{c^2}\right)\,,
\end{equation}
which could also be obtained by projecting the LL Eq.~(\ref{eq:LL}) with $u_{\mathcal{B}\mu}$ and using Eq.~(\ref{eq:dotr}). 

This differential equation can be readily integrated
\begin{equation}\label{eq:LLsol}
 \kappa(\tau) = \frac{\kappa_0 e^{-\Gamma\tau}}{1 +\kappa_0(1-e^{-\Gamma\tau})/c^2}\,,
\end{equation}
where the decay coefficient
\begin{equation}\label{eq:decay}
 \Gamma \equiv \frac{4\tau_0 q^2 }{\varepsilon_0 m^2}\mathcal{M}>0\,.
\end{equation}
The part of the four-velocity outside the $\mathcal{A}$ plane decays at a rate set by the invariant mass density of the field.

Equation~(\ref{eq:dotkappa}) shows that $\kappa$ decreases monotonically to zero for every initial condition. Then \req{eq:EuSPnonzero} implies that in this case $\mathcal{E}_u \rightarrow \mathcal{M}c^2$ and therefore $\Phi_u = c\sqrt{\mathcal{E}_u^2 - \mathcal{M}^2c^4}\to0$. The flux-free states are exactly the four-velocities that lie in the $\mathcal A$ plane,
$u^\mu=u^\mu_{\mathcal A}$. For such four-velocities $(F^2)^\mu_\nu u^\nu=\mathcal A^2u^\mu$ and
\begin{equation}
T^\mu_\nu u^\nu=c^2\varepsilon_0(\mathcal A^2+S)\,u^\mu
=\mathcal{M}c^2u^\mu,
\end{equation}
so $u^\mu$ is an eigenvector of the energy--momentum tensor.

Let us show that the $\mathcal A$ plane is invariant under the dynamics. Since $F^\mu_\nu$ commutes
with $(F^2)^\mu_\nu$, it commutes with the projectors Eqs.~(\ref{eq:projA},\ref{eq:projB}), and the Lorentz
force maps the $\mathcal A$ plane into itself because
\begin{equation}
 F^\mu_\nu u_\mathcal{A}^\nu = F^\mu_\nu (\Pi_\mathcal{A})^\nu_\alpha u^\alpha = (\Pi_\mathcal{A})^\mu_\nu F^\nu_\alpha u^\alpha \in \mathcal{A}\,.
\end{equation}
Moreover, for
$u^\mu=u_{\mathcal A}^\mu$ the radiation-reaction
term in \req{eq:LL} vanishes identically,
\begin{equation}
\tau_0 \frac{q^2}{m^2}P^\mu_\nu (F^2)^\nu_\alpha u^\alpha = \tau_0 \frac{q^2\mathcal{A}^2}{m^2}P^\mu_\nu u^\nu = 0 .
\end{equation}
Thus, once $u_{\mathcal B}^\mu=0$ it stays zero. The motion is then RR-free, and it obeys the Lorentz force $a^\mu=(q/m)F^\mu_\nu u^\nu$. Differentiating,
\begin{equation}
\dot a^\mu=\frac{q}{m}F^\mu_\nu a^\nu=\frac{q^2}{m^2}(F^2)^\mu_\nu u^\nu
=\frac{q^2\mathcal A^2}{m^2}\,u^\mu\,.
\end{equation}
The invariant magnitude of four-acceleration is 
\begin{equation}
\qquad a^2=-\frac{q^2}{m^2}\,u\cdot F^2\cdot u=-\frac{q^2\mathcal A^2c^2}{m^2}.
\end{equation}
If $\mathcal{A} \neq 0$ then $\dot{a}^\mu \propto u^\mu$, which is Rindler's RR-free motion with constant proper acceleration $|q|\mathcal{A}c/m$ \cite{Rindler:1960zz}. If $\mathcal{A}=0$ then $a^2=0$ and $a\cdot u=0$ with both
vectors in the two-dimensional $\mathcal A$ plane. Any four-vector orthogonal to the timelike $u^\mu$ is either spacelike ($a^2 < 0$) or zero, so
$a^\mu=0$. This means that the motion is force-free and uniform. In both cases the approach to the attractor is governed by the decay of $\kappa = |u_{\mathcal B}^2|$, Eq.~(\ref{eq:LLsol}). The discussion of how the particle trajectories in specific field configurations converge to these two cases is included in Sec.~\ref{sec:examples}.

\section{Crossed Fields}\label{sec:cross}
\subsection{Motion in null fields \texorpdfstring{$\bm{S^2+P^2= 0}$}{}}\label{sec:null}
In the case of $\mathcal{A}^2 + \mathcal{B}^2 = 2\sqrt{S^2 + P^2} = 0$ the decomposition introduced in Sec.~{\ref{sec:nonnull}} is not possible. This is the case of crossed fields of equal magnitude and it is necessary to introduce an additional lightlike four-vector 
\begin{equation}\label{eq:kmu}
 k^\mu = \frac{E}{c}(1, \hat{\boldsymbol{E}}\times\hat{\boldsymbol{B}})\,, \quad k^2 = 0\,,
\end{equation}
which defines the direction perpendicular to the plane spanned by the fields. In this case, the EM tensor can be written as
\begin{equation}
 F^{\mu\nu} = \varepsilon^\mu k^\nu - \varepsilon^\nu k^\mu\,,\quad (F^2)^\mu_\nu = k^\mu k_\nu\,,
\end{equation}
where $\varepsilon^\mu = (0, \hat{\boldsymbol{E}})$ is the direction of the electric field satisfying $\varepsilon^2=-1$, $k\cdot \varepsilon = 0$. The effect of RR can be studied using the invariant product $k \cdot u$ which, without RR, is conserved in constant fields. The EM energy-momentum tensor in this field is
\begin{equation}
 T^\mu_\nu = c^2 \varepsilon_0 k^\mu k_\nu\,,
\end{equation}
so that the invariant mass density of the field vanishes
\begin{equation}
 \mathcal{M}(S^2+P^2=0) = \frac{1}{c^3}\sqrt{u \cdot T^2 \cdot u} = 0\,.
\end{equation}
The energy density seen by the moving charge is
\begin{equation}\label{eq:Eunull}
 \mathcal{E}_u(S^2+P^2=0) = \varepsilon_0 (k \cdot u)^2
\end{equation}
and substituting both into Eq. (\ref{eq:energydensityrelax}) yields
\begin{equation}\label{eq:kcdotu}
 \frac{d}{d\tau}[k \cdot u(\tau)] = -\tau_0 \frac{q^2}{m^2 c^2}[k\cdot u(\tau)]^3\,,
\end{equation}
which is a known equation obtainable by projecting the LL equation of motion Eq.~(\ref{eq:LL}) on the $k_\mu$ direction. It has an analytic solution that matches the infinite wavelength limit of the plane-wave solution~\cite{Piazza:2008yvh,Hadad:2010mt}
\begin{equation}
 [k\cdot u(\tau)]^2 = \frac{[k\cdot u(0)]^2}{1 +2 [k \cdot u(0)]^2 \tau_0 \frac{q^2}{m^2c^2}\tau }\,.
\end{equation}
As $\tau$ grows, the invariant $k\cdot u$ goes to zero, which means that the four-velocity $u^\mu$ is asymptotically approaching the null direction $k^\mu$, since the only four-vectors orthogonal to $k^\mu$ in addition to $k^\mu$ itself are spacelike. The magnitude of the Lorentz force acceleration depends on the invariant $k\cdot u$ as
\begin{equation}
 a_\text{LF}^2(\tau) = -\frac{q^2}{m^2} [k \cdot u(\tau)]^2
\end{equation}
and also decreases. Therefore, even motion in the null field tends to the RR-free case due to the RR effect. In this case the RR-free motion is approached but never reached in the limit of the motion becoming uniform and lightlike.

\subsection{Ultrarelativistic Motion}\label{sec:RelMotion}
For an ultrarelativistic charge, the null field case effectively applies. A fast-moving charge sees in its rest frame almost crossed fields of nearly equal electric and magnetic magnitudes ($E\approx Bc$). Together with the requirement on slowly varying fields, this is known in strong-field QED as the locally constant, crossed field
approximation (LCFA)~\cite{Ritus:1985vta,DiPiazza:2011tq,Gonoskov:2021hwf}. In this regime $\mathcal{E}_u \gg \mathcal{M}c^2$ and the energy density seen by the charge obeys Eq.~(\ref{eq:energydensityrelax}) with $\mathcal{M}^2c^4$ neglected so that it evolves as
\begin{equation}\label{eq:LCFA}
 \mathcal{E}_u(\tau) \approx \frac{\mathcal{E}_u(0)}{1 + \frac{2\tau_0 q^2}{\varepsilon_0 m^2c^2}\mathcal{E}_u(0)\tau}\,.
\end{equation}
Within LCFA validity, this decay is universal because it does not depend on the field configuration or on the initial direction of motion. The approximation breaks down when $\mathcal{E}_u$ has decayed to the order of $\mathcal{M}c^2$. 

Notably, the same quantity controls the onset of quantum effects. The quantum nonlinearity parameter of the moving charge is~\cite{Ritus:1985vta}
\begin{equation}
 \chi = \frac{\hbar |q|}{m^3 c^3}\sqrt{-(F^\mu_\nu p^\nu)^2} = \frac{\hbar |q|}{m^2 c^3}\sqrt{\frac{\mathcal{E}_u}{\varepsilon_0}-S c^2}\,,
\end{equation}
where $p^\mu = m u^\mu$ is the particle four-momentum and the definition of $T^\mu_\nu$ \req{eq:Tmunu} was substituted. Within LCFA $\varepsilon_0 c^2|S| \ll \mathcal{E}_u$ and the expression further simplifies
\begin{equation}
 \chi_\text{LCFA}^2 \approx \frac{\hbar^2 q^2}{\varepsilon_0 m^4 c^6}\mathcal{E}_u\,,
\end{equation}
where $\mathcal{E}_u$ follows Eq.~(\ref{eq:LCFA}). The classical treatment presented in this letter requires $\chi \ll 1$. For $\chi \gtrsim 1$ the dynamical evolution is a quantum stochastic process due to the probabilistic nature of photon emission until the classical limit is reached.

\section{Attractor trajectories in \texorpdfstring{$\bm{S^2+P^2\neq 0}$}{} fields}\label{sec:examples}
The solutions for constant fields are well known, see e.g.~\cite{Yaremko:2013eqe}. Here they are discussed in the limit of long proper times using the insights introduced in this work. Starting with the most general case of $P \neq 0$ where there exists a Lorentz frame in which $\boldsymbol{E}\parallel \boldsymbol{B}$. The invariant mass density of the field is given by Eq.~(\ref{eq:ESPnonzero}) as $\mathcal{M} = \varepsilon_0 (B^2 + E^2/c^2)/2$ and the decay coefficient of Eq.~(\ref{eq:decay}) becomes
\begin{equation}\label{eq:gammagen}
    \Gamma = 2 \tau_0(\omega_E^2 + \omega_B^2)\,, \quad \omega_B \equiv \frac{|q| B}{m}\,, \quad \omega_E \equiv \frac{|q|E}{mc}\,,
\end{equation}
where $\omega_B$ is the proper-time cyclotron frequency and $c \omega_E$ is the proper acceleration of Rindler motion in this field. Let us work in the frame where the common direction of both fields is oriented along the $\hat{z}$ direction. The four-velocity is then $u^\mu = \gamma(c, \boldsymbol{v}_\perp,v_\parallel)$ with $\gamma$ being the Lorentz factor $\gamma = (1-v^2/c^2)^{-1/2}$ and $\boldsymbol{v}_\perp$, $v_\parallel$ three-velocities perpendicular and parallel to the $\hat{z}$ direction respectively. Using the definitions Eqs.~(\ref{eq:projA},\ref{eq:projB}) the projectors are $(\Pi_\mathcal{A})^\mu_\nu=\mathrm{diag}(1,0,0,1)$ and $(\Pi_\mathcal{B})^\mu_\nu=\mathrm{diag}(0,1,1,0)$. The four-velocity is then decomposed as
\begin{equation}
 u_\mathcal{A}^\mu =\gamma(c,0,0,v_\parallel)\,, \quad u_\mathcal{B}^\mu =\gamma(0,\boldsymbol{v}_\perp,0)\,.
\end{equation}
According to Eq.~(\ref{eq:uB2})
\begin{equation}\label{eq:kappagen}
 \kappa = - u_\mathcal{B}^2 = (\gamma v_\perp)^2 = \frac{p_\perp^2}{m^2}\,.
\end{equation}
Therefore the $\kappa$ variable is proportional to the square of the magnitude of transverse momentum. It decays following Eq.~(\ref{eq:LLsol}). Moreover, it is possible to Lorentz transform along the $\hat{z}$ direction to an instantaneous frame where $v_\parallel = 0$ and the perpendicular velocity satisfies
\begin{equation}
 \frac{1}{v_\perp'^2}\frac{dv_\perp'^2}{d\tau} = -\Gamma\,,
\end{equation}
in this frame the decay coefficient $\Gamma$ applies to the square of the magnitude of the transverse part of the three-velocity directly. In general, the trajectory gyrates around the magnetic field direction with decreasing perpendicular velocity and is simultaneously accelerated by the electric field along the same direction. The asymptotic trajectory is the Rindler motion along the $\hat{z}$ direction. 

\subsection{Special cases for \texorpdfstring{$\bm{P=0}$}{}}
If we set $P = 0$ the general example breaks into two special cases:

\textbf{i) Purely magnetic case} ($P=0$, $S > 0$) for which in some Lorentz frame a pure magnetic field $\boldsymbol{B}$ exists. The decay coefficient \req{eq:gammagen} $\Gamma = 2\tau_0 \omega_B^2$ is the well-known cyclotron damping coefficient~\cite{Yaremko:2013eqe}. The particle gyrates around the magnetic field direction with decreasing transverse momentum as it asymptotically approaches inertial motion along this axis.

\textbf{ii) Purely electric case} ($P = 0$, $S < 0$) allowing in some Lorentz frame a pure electric field $\boldsymbol{E}$. The decay coefficient for the transverse momentum is according to \req{eq:gammagen} $\Gamma = 2\tau_0 \omega_E^2$. Unlike in the magnetic case, the asymptotic motion is not inertial, because the charge keeps accelerating along $\boldsymbol{E}$. Without RR the transverse momentum would be conserved but now it decays and the particle approaches the RR-free Rindler trajectory fixed to a straight line. The radiated energy originates from the change of the particle speed along the converged trajectory.

\subsection{Evolution towards RR-free trajectory}\label{sec:evol}
\begin{figure}
 \includegraphics[width=\columnwidth]{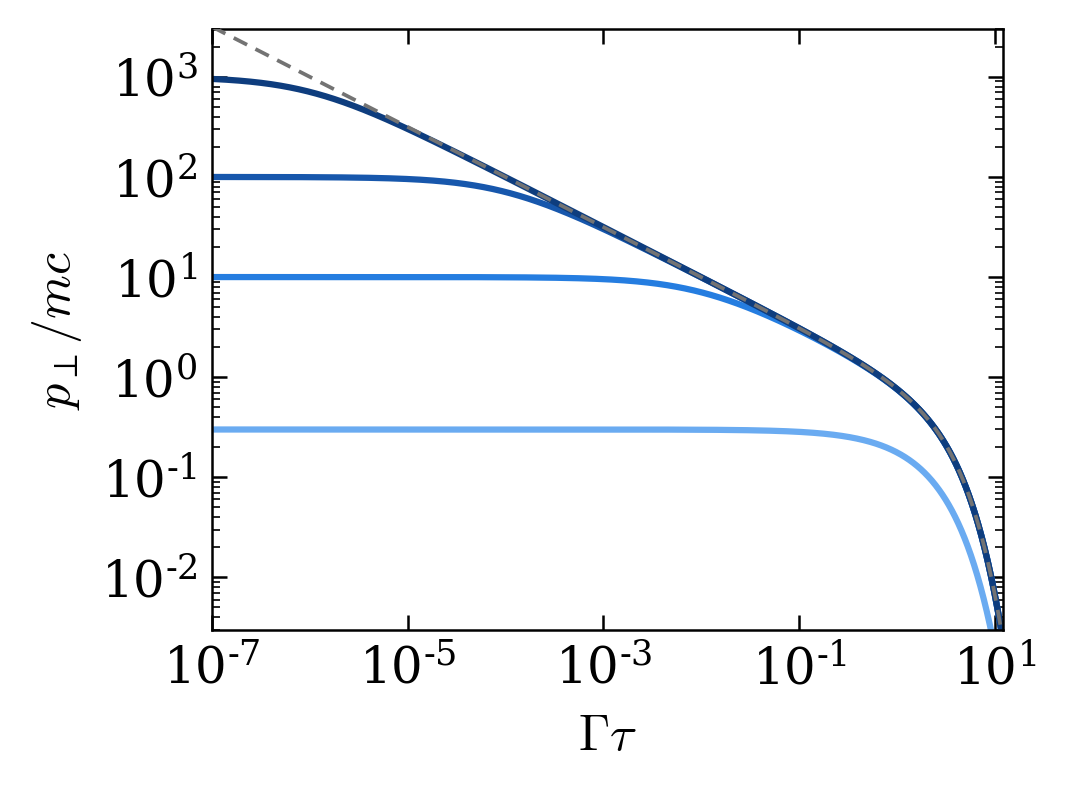}
 \caption{Decay of the transverse momentum \req{eq:trans_mom} as a function of proper time in units of $1/\Gamma$, for initial values $p_\perp/mc=0.3,\,10,\,100,\,1000$ (light to dark solid lines). Dashed line is the limit $\kappa(0)\to\infty$.}
 \label{fig:transverse_decay}
\end{figure}

In all situations the decay of the transverse momentum can be plotted as the same function, see Eq.~(\ref{eq:kappagen})
\begin{equation}\label{eq:trans_mom}
 \frac{p_\perp(\tau)}{mc} = \frac{\sqrt{\kappa(\tau)}}{c}\,.
\end{equation}
The normalized transverse momentum depends on its initial value and evolves following \req{eq:LLsol} as a function of proper time, rescaled by the corresponding decay rate $\Gamma$, see Eq.~(\ref{eq:gammagen}). Figure~\ref{fig:transverse_decay} shows that a large initial transverse momentum with respect to asymptotic motion is lost quickly, because the decay rate in \req{eq:dotkappa} is enhanced while $\kappa \gg c^2$. When $\kappa(0)\gg c^2$ the dependence on the initial momentum is lost after $\Gamma\tau\sim c^2/\kappa(0)$, followed by the algebraic decay $p_\perp/mc\simeq (\Gamma\tau)^{-1/2}$ and, for $\Gamma\tau\gtrsim 1$, the exponential decay $p_\perp/mc \propto e^{-\Gamma\tau/2}$.

This means that the RR-free motion is approximately achieved in finite proper time as the dashed bound in Fig.~\ref{fig:transverse_decay} shows. 

\section{Conclusions and discussion}\label{sec:discussion}
A rather simple consideration of the particle-observed energy density $\mathcal{E}_u$, \req{eq:Eu}, and invariant mass density $\mathcal{M}c^2$, \req{eq:E},  of the external electromagnetic Maxwellian force-field demonstrates the existence of attractor trajectories towards which particle dynamics evolves when subject to vacuum friction. We studied the analytically soluble example of a classical charge in a quasi-constant and quasi-homogeneous electromagnetic external force-field while describing the vacuum friction effect using the LL reaction force. We have shown that the energy flux, \req{eq:Eu}, of the external field seen by the charge decays monotonically to zero, with logarithmic rate proportional to $\mathcal{E}_u$. For flux-free motion the RR force vanishes identically. Every trajectory in a non-null field is driven towards Rindler motion, unless the field is purely magnetic, in which case the attractor trajectory is force-free motion. In a crossed null field the proper acceleration and the RR force decay, and the velocity tends to $c$ along the null direction of the field while the energy keeps growing.

In view of prior work~\cite{Bulanov2011,Gonoskov:2017lyz,Samsonov:2018skj,Ekman:2021vwg,Cai:2022mkw,Cai:2023ykg,Samsonov2023}, we believe that our result is one example of a more general mechanism that allows further analytical insights. For example, generalization to arbitrary fields using LL reaction force is obtained by differentiating~\req{eq:Eu} along a worldline leading to the generalized form of~\req{eq:introEvol}
\begin{equation}\label{eq:Eu_general}
\begin{split}
\dot{\mathcal{E}}_u =
&-\frac{2\tau_0 q^2}{\varepsilon_0 m^2 c^2}\left(\mathcal{E}_u^2-\mathcal{M}^2c^4\right)
+\frac{1}{c^2}u \cdot \dot{T}\cdot u\\
&+\frac{2\tau_0 q}{m c^2}\, u\cdot T\cdot \dot{F}\cdot u\,,
\end{split}
\end{equation}
where $\mathcal{M}$ is evaluated at the position of the charge. The second term arises due to the change in the field energy-momentum tensor $T$ as the particle moves through the inhomogeneous and/or time-dependent field. The third term comes from the derivative term of the LL force.

This shows that RR drives the charge towards the locally flux-free state whenever the first term dominates the other two. In this regime the field can be treated as locally constant, and the charge follows the instantaneous flux-free direction. This gives a covariant criterion for evolution towards the RR-free direction, a process which was simulated numerically in strong fields~\cite{Gonoskov:2017lyz,Samsonov:2018skj,Cai:2022mkw,Cai:2023ykg,Samsonov2023}. Using \req{eq:introEvol}, or more generally \req{eq:Eu_general}, the dynamical evolution of particle motion to RR-free states can be directly monitored in particle-in-cell simulations and used to study plasma behavior in external fields \cite{lindl1995development,craxton2015direct}. 

The framework developed here is not specific to the
LL vacuum friction as the dynamics in \req{eq:dflux} can be generalized to consider other vacuum friction models. However, it is clear that any analytical solution will be made complicated by: i) the nonlinear (LAD) or implicit (EFO) dependence on acceleration and ii) its derivative (LAD). To first order in external force the behavior of all models is identical~\cite{Price:2021zqq} and it is known that Rindler motion and inertial motion exist as RR-free trajectories in both EFO and LAD models to all orders. Because the EFO reaction force contains no higher-order derivatives, \req{eq:dflux} is more likely to remain analytically solvable in that case. So far, we have not considered other interesting classical vacuum friction models such as the discretized approach~\cite{caldirola1956new} or warping approach~\cite{Formanek:2020zwc}. Nor did we try to address any quantum analog given that our argument relies on particle trajectories parametrized by proper time, both of which are conceptually absent in quantum theory.

\acknowledgments
J.R. acknowledges support from the Arizona Board of Regents through the Technology and Research Initiative Fund (TRIF) Opportunity Funds for Fusion Energy (Grant 5852739).


\begin{thebibliography}{45}%
	\makeatletter
	\providecommand \@ifxundefined [1]{%
		\@ifx{#1\undefined}
	}%
	\providecommand \@ifnum [1]{%
		\ifnum #1\expandafter \@firstoftwo
		\else \expandafter \@secondoftwo
		\fi
	}%
	\providecommand \@ifx [1]{%
		\ifx #1\expandafter \@firstoftwo
		\else \expandafter \@secondoftwo
		\fi
	}%
	\providecommand \natexlab [1]{#1}%
	\providecommand \enquote  [1]{``#1''}%
	\providecommand \bibnamefont  [1]{#1}%
	\providecommand \bibfnamefont [1]{#1}%
	\providecommand \citenamefont [1]{#1}%
	\providecommand \href@noop [0]{\@secondoftwo}%
	\providecommand \href [0]{\begingroup \@sanitize@url \@href}%
	\providecommand \@href[1]{\@@startlink{#1}\@@href}%
	\providecommand \@@href[1]{\endgroup#1\@@endlink}%
	\providecommand \@sanitize@url [0]{\catcode `\\12\catcode `\$12\catcode
		`\&12\catcode `\#12\catcode `\^12\catcode `\_12\catcode `\%12\relax}%
	\providecommand \@@startlink[1]{}%
	\providecommand \@@endlink[0]{}%
	\providecommand \url  [0]{\begingroup\@sanitize@url \@url }%
	\providecommand \@url [1]{\endgroup\@href {#1}{\urlprefix }}%
	\providecommand \urlprefix  [0]{URL }%
	\providecommand \Eprint [0]{\href }%
	\providecommand \doibase [0]{https://doi.org/}%
	\providecommand \selectlanguage [0]{\@gobble}%
	\providecommand \bibinfo  [0]{\@secondoftwo}%
	\providecommand \bibfield  [0]{\@secondoftwo}%
	\providecommand \translation [1]{[#1]}%
	\providecommand \BibitemOpen [0]{}%
	\providecommand \bibitemStop [0]{}%
	\providecommand \bibitemNoStop [0]{.\EOS\space}%
	\providecommand \EOS [0]{\spacefactor3000\relax}%
	\providecommand \BibitemShut  [1]{\csname bibitem#1\endcsname}%
	\let\auto@bib@innerbib\@empty
	\bibitem [{\citenamefont {Steinmetz}\ \emph {et~al.}(2027)\citenamefont
		{Steinmetz}, \citenamefont {Labun}, \citenamefont {Formanek}, \citenamefont
		{Price}, \citenamefont {Grayson},\ and\ \citenamefont
		{Rafelski}}]{Steinmetz2027}%
	\BibitemOpen
	\bibfield  {author} {\bibinfo {author} {\bibfnamefont {A.~J.}\ \bibnamefont
			{Steinmetz}}, \bibinfo {author} {\bibfnamefont {L.}~\bibnamefont {Labun}},
		\bibinfo {author} {\bibfnamefont {M.~S.}\ \bibnamefont {Formanek}}, \bibinfo
		{author} {\bibfnamefont {W.}~\bibnamefont {Price}}, \bibinfo {author}
		{\bibfnamefont {C.~M.}\ \bibnamefont {Grayson}},\ and\ \bibinfo {author}
		{\bibfnamefont {J.}~\bibnamefont {Rafelski}},\ }\bibfield  {title} {\bibinfo
		{title} {Classical and quantum particle dynamics and radiation in strong
			fields},\ }\href@noop {} {\bibfield  {journal} {\bibinfo  {journal} {{\it in
					preparation for EPJ-C, review section}}\ } (\bibinfo {year}
		{2027})}\BibitemShut {NoStop}%
	\bibitem [{\citenamefont {Dirac}(1938)}]{Dirac:1938nz}%
	\BibitemOpen
	\bibfield  {author} {\bibinfo {author} {\bibfnamefont {P.~A.~M.}\
			\bibnamefont {Dirac}},\ }\bibfield  {title} {\bibinfo {title} {{Classical
				theory of radiating electrons}},\ }\href
	{https://doi.org/10.1098/rspa.1938.0124} {\bibfield  {journal} {\bibinfo
			{journal} {Proc. Roy. Soc. Lond. A}\ }\textbf {\bibinfo {volume} {167}},\
		\bibinfo {pages} {148} (\bibinfo {year} {1938})}\BibitemShut {NoStop}%
	\bibitem [{\citenamefont {Landau}\ and\ \citenamefont
		{Lifshitz}(1975)}]{Landau:1975pou}%
	\BibitemOpen
	\bibfield  {author} {\bibinfo {author} {\bibfnamefont {L.~D.}\ \bibnamefont
			{Landau}}\ and\ \bibinfo {author} {\bibfnamefont {E.~M.}\ \bibnamefont
			{Lifshitz}},\ }\href {https://doi.org/10.1016/c2009-0-14608-1} {\emph
		{\bibinfo {title} {{The Classical Theory of Fields}}}},\ \bibinfo {series}
	{Course of Theoretical Physics}, Vol.~\bibinfo {volume} {2}\ (\bibinfo
	{publisher} {Pergamon Press},\ \bibinfo {address} {Oxford},\ \bibinfo {year}
	{1975})\BibitemShut {NoStop}%
	\bibitem [{\citenamefont {Eliezer}(1948)}]{Eliezer:1948qkl}%
	\BibitemOpen
	\bibfield  {author} {\bibinfo {author} {\bibfnamefont {C.~J.}\ \bibnamefont
			{Eliezer}},\ }\bibfield  {title} {\bibinfo {title} {{On the classical theory
				of particles}},\ }\href {https://doi.org/10.1098/rspa.1948.0096} {\bibfield
		{journal} {\bibinfo  {journal} {Proc. Roy. Soc. Lond. A}\ }\textbf {\bibinfo
			{volume} {194}},\ \bibinfo {pages} {543} (\bibinfo {year}
		{1948})}\BibitemShut {NoStop}%
	\bibitem [{\citenamefont {Ford}\ and\ \citenamefont
		{O'Connell}(1991)}]{Ford:1991ntp}%
	\BibitemOpen
	\bibfield  {author} {\bibinfo {author} {\bibfnamefont {G.~W.}\ \bibnamefont
			{Ford}}\ and\ \bibinfo {author} {\bibfnamefont {R.~F.}\ \bibnamefont
			{O'Connell}},\ }\bibfield  {title} {\bibinfo {title} {{Radiation reaction in
				electrodynamics and the elimination of runaway solutions}},\ }\href
	{https://doi.org/10.1016/0375-9601(91)90054-C} {\bibfield  {journal}
		{\bibinfo  {journal} {Phys. Lett. A}\ }\textbf {\bibinfo {volume} {157}},\
		\bibinfo {pages} {217} (\bibinfo {year} {1991})}\BibitemShut {NoStop}%
	\bibitem [{\citenamefont {Price}\ \emph {et~al.}(2022)\citenamefont {Price},
		\citenamefont {Formanek},\ and\ \citenamefont {Rafelski}}]{Price:2021zqq}%
	\BibitemOpen
	\bibfield  {author} {\bibinfo {author} {\bibfnamefont {W.}~\bibnamefont
			{Price}}, \bibinfo {author} {\bibfnamefont {M.}~\bibnamefont {Formanek}},\
		and\ \bibinfo {author} {\bibfnamefont {J.}~\bibnamefont {Rafelski}},\
	}\bibfield  {title} {\bibinfo {title} {{Radiation reaction and limiting
				acceleration}},\ }\href {https://doi.org/10.1103/PhysRevD.105.016024}
	{\bibfield  {journal} {\bibinfo  {journal} {Phys. Rev. D}\ }\textbf {\bibinfo
			{volume} {105}},\ \bibinfo {pages} {016024} (\bibinfo {year}
		{2022})}\BibitemShut {NoStop}%
	\bibitem [{\citenamefont {Di~Piazza}\ \emph {et~al.}(2012)\citenamefont
		{Di~Piazza}, \citenamefont {M{\"u}ller}, \citenamefont {Hatsagortsyan},\ and\
		\citenamefont {Keitel}}]{DiPiazza:2011tq}%
	\BibitemOpen
	\bibfield  {author} {\bibinfo {author} {\bibfnamefont {A.}~\bibnamefont
			{Di~Piazza}}, \bibinfo {author} {\bibfnamefont {C.}~\bibnamefont
			{M{\"u}ller}}, \bibinfo {author} {\bibfnamefont {K.~Z.}\ \bibnamefont
			{Hatsagortsyan}},\ and\ \bibinfo {author} {\bibfnamefont {C.~H.}\
			\bibnamefont {Keitel}},\ }\bibfield  {title} {\bibinfo {title} {{Extremely
				high-intensity laser interactions with fundamental quantum systems}},\ }\href
	{https://doi.org/10.1103/RevModPhys.84.1177} {\bibfield  {journal} {\bibinfo
			{journal} {Rev. Mod. Phys.}\ }\textbf {\bibinfo {volume} {84}},\ \bibinfo
		{pages} {1177} (\bibinfo {year} {2012})}\BibitemShut {NoStop}%
	\bibitem [{\citenamefont {Blackburn}(2020)}]{Blackburn:2019rfv}%
	\BibitemOpen
	\bibfield  {author} {\bibinfo {author} {\bibfnamefont {T.~G.}\ \bibnamefont
			{Blackburn}},\ }\bibfield  {title} {\bibinfo {title} {{Radiation reaction in
				electron-beam interactions with high-intensity lasers}},\ }\href
	{https://doi.org/10.1007/s41614-020-0042-0} {\bibfield  {journal} {\bibinfo
			{journal} {Rev. Mod. Plasma Phys.}\ }\textbf {\bibinfo {volume} {4}},\
		\bibinfo {pages} {5} (\bibinfo {year} {2020})}\BibitemShut {NoStop}%
	\bibitem [{\citenamefont {Gonoskov}\ \emph {et~al.}(2022)\citenamefont
		{Gonoskov}, \citenamefont {Blackburn}, \citenamefont {Marklund},\ and\
		\citenamefont {Bulanov}}]{Gonoskov:2021hwf}%
	\BibitemOpen
	\bibfield  {author} {\bibinfo {author} {\bibfnamefont {A.}~\bibnamefont
			{Gonoskov}}, \bibinfo {author} {\bibfnamefont {T.~G.}\ \bibnamefont
			{Blackburn}}, \bibinfo {author} {\bibfnamefont {M.}~\bibnamefont
			{Marklund}},\ and\ \bibinfo {author} {\bibfnamefont {S.~S.}\ \bibnamefont
			{Bulanov}},\ }\bibfield  {title} {\bibinfo {title} {{Charged particle motion
				and radiation in strong electromagnetic fields}},\ }\href
	{https://doi.org/10.1103/RevModPhys.94.045001} {\bibfield  {journal}
		{\bibinfo  {journal} {Rev. Mod. Phys.}\ }\textbf {\bibinfo {volume} {94}},\
		\bibinfo {pages} {045001} (\bibinfo {year} {2022})}\BibitemShut {NoStop}%
	\bibitem [{\citenamefont {Fedotov}\ \emph {et~al.}(2023)\citenamefont
		{Fedotov}, \citenamefont {Ilderton}, \citenamefont {Karbstein}, \citenamefont
		{King}, \citenamefont {Seipt}, \citenamefont {Taya},\ and\ \citenamefont
		{Torgrimsson}}]{Fedotov:2022ely}%
	\BibitemOpen
	\bibfield  {author} {\bibinfo {author} {\bibfnamefont {A.}~\bibnamefont
			{Fedotov}}, \bibinfo {author} {\bibfnamefont {A.}~\bibnamefont {Ilderton}},
		\bibinfo {author} {\bibfnamefont {F.}~\bibnamefont {Karbstein}}, \bibinfo
		{author} {\bibfnamefont {B.}~\bibnamefont {King}}, \bibinfo {author}
		{\bibfnamefont {D.}~\bibnamefont {Seipt}}, \bibinfo {author} {\bibfnamefont
			{H.}~\bibnamefont {Taya}},\ and\ \bibinfo {author} {\bibfnamefont
			{G.}~\bibnamefont {Torgrimsson}},\ }\bibfield  {title} {\bibinfo {title}
		{{Advances in QED with intense background fields}},\ }\href
	{https://doi.org/10.1016/j.physrep.2023.01.003} {\bibfield  {journal}
		{\bibinfo  {journal} {Phys. Rept.}\ }\textbf {\bibinfo {volume} {1010}},\
		\bibinfo {pages} {1} (\bibinfo {year} {2023})}\BibitemShut {NoStop}%
	\bibitem [{\citenamefont {Poder}\ \emph {et~al.}(2018)\citenamefont {Poder},
		\citenamefont {Tamburini}, \citenamefont {Sarri}, \citenamefont {Di~Piazza},
		\citenamefont {Kuschel}, \citenamefont {Baird}, \citenamefont {Behm},
		\citenamefont {Bohlen}, \citenamefont {Cole}, \citenamefont {Corvan} \emph
		{et~al.}}]{Poder:2017dpw}%
	\BibitemOpen
	\bibfield  {author} {\bibinfo {author} {\bibfnamefont {K.}~\bibnamefont
			{Poder}}, \bibinfo {author} {\bibfnamefont {M.}~\bibnamefont {Tamburini}},
		\bibinfo {author} {\bibfnamefont {G.}~\bibnamefont {Sarri}}, \bibinfo
		{author} {\bibfnamefont {A.}~\bibnamefont {Di~Piazza}}, \bibinfo {author}
		{\bibfnamefont {S.}~\bibnamefont {Kuschel}}, \bibinfo {author} {\bibfnamefont
			{C.~D.}\ \bibnamefont {Baird}}, \bibinfo {author} {\bibfnamefont
			{K.}~\bibnamefont {Behm}}, \bibinfo {author} {\bibfnamefont {S.}~\bibnamefont
			{Bohlen}}, \bibinfo {author} {\bibfnamefont {J.~M.}\ \bibnamefont {Cole}},
		\bibinfo {author} {\bibfnamefont {D.~J.}\ \bibnamefont {Corvan}}, \emph
		{et~al.},\ }\bibfield  {title} {\bibinfo {title} {{Experimental Signatures of
				the Quantum Nature of Radiation Reaction in the Field of an Ultraintense
				Laser}},\ }\href {https://doi.org/10.1103/PhysRevX.8.031004} {\bibfield
		{journal} {\bibinfo  {journal} {Phys. Rev. X}\ }\textbf {\bibinfo {volume}
			{8}},\ \bibinfo {pages} {031004} (\bibinfo {year} {2018})}\BibitemShut
	{NoStop}%
	\bibitem [{\citenamefont {Cole}\ \emph {et~al.}(2018)\citenamefont {Cole},
		\citenamefont {Behm}, \citenamefont {Gerstmayr}, \citenamefont {Blackburn},
		\citenamefont {Wood}, \citenamefont {Baird}, \citenamefont {Duff},
		\citenamefont {Harvey}, \citenamefont {Ilderton}, \citenamefont {Joglekar}
		\emph {et~al.}}]{Cole:2017zca}%
	\BibitemOpen
	\bibfield  {author} {\bibinfo {author} {\bibfnamefont {J.~M.}\ \bibnamefont
			{Cole}}, \bibinfo {author} {\bibfnamefont {K.~T.}\ \bibnamefont {Behm}},
		\bibinfo {author} {\bibfnamefont {E.}~\bibnamefont {Gerstmayr}}, \bibinfo
		{author} {\bibfnamefont {T.~G.}\ \bibnamefont {Blackburn}}, \bibinfo {author}
		{\bibfnamefont {J.~C.}\ \bibnamefont {Wood}}, \bibinfo {author}
		{\bibfnamefont {C.~D.}\ \bibnamefont {Baird}}, \bibinfo {author}
		{\bibfnamefont {M.~J.}\ \bibnamefont {Duff}}, \bibinfo {author}
		{\bibfnamefont {C.}~\bibnamefont {Harvey}}, \bibinfo {author} {\bibfnamefont
			{A.}~\bibnamefont {Ilderton}}, \bibinfo {author} {\bibfnamefont {A.~S.}\
			\bibnamefont {Joglekar}}, \emph {et~al.},\ }\bibfield  {title} {\bibinfo
		{title} {{Experimental evidence of radiation reaction in the collision of a
				high-intensity laser pulse with a laser-wakefield accelerated electron
				beam}},\ }\href {https://doi.org/10.1103/PhysRevX.8.011020} {\bibfield
		{journal} {\bibinfo  {journal} {Phys. Rev. X}\ }\textbf {\bibinfo {volume}
			{8}},\ \bibinfo {pages} {011020} (\bibinfo {year} {2018})}\BibitemShut
	{NoStop}%
	\bibitem [{\citenamefont {Mirzaie}\ \emph {et~al.}(2024)\citenamefont
		{Mirzaie}, \citenamefont {Hojbota}, \citenamefont {Kim}, \citenamefont
		{Pathak}, \citenamefont {Pak}, \citenamefont {Kim}, \citenamefont {Lee},
		\citenamefont {Yoon}, \citenamefont {Lee}, \citenamefont {Rhee} \emph
		{et~al.}}]{Mirzaie:2024iey}%
	\BibitemOpen
	\bibfield  {author} {\bibinfo {author} {\bibfnamefont {M.}~\bibnamefont
			{Mirzaie}}, \bibinfo {author} {\bibfnamefont {C.~I.}\ \bibnamefont
			{Hojbota}}, \bibinfo {author} {\bibfnamefont {D.~Y.}\ \bibnamefont {Kim}},
		\bibinfo {author} {\bibfnamefont {V.~B.}\ \bibnamefont {Pathak}}, \bibinfo
		{author} {\bibfnamefont {T.~G.}\ \bibnamefont {Pak}}, \bibinfo {author}
		{\bibfnamefont {C.~M.}\ \bibnamefont {Kim}}, \bibinfo {author} {\bibfnamefont
			{H.~W.}\ \bibnamefont {Lee}}, \bibinfo {author} {\bibfnamefont {J.~W.}\
			\bibnamefont {Yoon}}, \bibinfo {author} {\bibfnamefont {S.~K.}\ \bibnamefont
			{Lee}}, \bibinfo {author} {\bibfnamefont {Y.~J.}\ \bibnamefont {Rhee}}, \emph
		{et~al.},\ }\bibfield  {title} {\bibinfo {title} {All-optical nonlinear
			{C}ompton scattering performed with a multi-petawatt laser},\ }\href
	{https://doi.org/10.1038/s41566-024-01550-8} {\bibfield  {journal} {\bibinfo
			{journal} {Nature Photon.}\ }\textbf {\bibinfo {volume} {18}},\ \bibinfo
		{pages} {1212} (\bibinfo {year} {2024})}\BibitemShut {NoStop}%
	\bibitem [{\citenamefont {Los}\ \emph {et~al.}(2026)\citenamefont {Los},
		\citenamefont {Gerstmayr}, \citenamefont {Arran}, \citenamefont {Streeter},
		\citenamefont {Colgan}, \citenamefont {Cobo}, \citenamefont {Kettle},
		\citenamefont {Blackburn}, \citenamefont {Bourgeois}, \citenamefont {Calvin}
		\emph {et~al.}}]{Los:2024ysw}%
	\BibitemOpen
	\bibfield  {author} {\bibinfo {author} {\bibfnamefont {E.~E.}\ \bibnamefont
			{Los}}, \bibinfo {author} {\bibfnamefont {E.}~\bibnamefont {Gerstmayr}},
		\bibinfo {author} {\bibfnamefont {C.}~\bibnamefont {Arran}}, \bibinfo
		{author} {\bibfnamefont {M.~J.~V.}\ \bibnamefont {Streeter}}, \bibinfo
		{author} {\bibfnamefont {C.}~\bibnamefont {Colgan}}, \bibinfo {author}
		{\bibfnamefont {C.~C.}\ \bibnamefont {Cobo}}, \bibinfo {author}
		{\bibfnamefont {B.}~\bibnamefont {Kettle}}, \bibinfo {author} {\bibfnamefont
			{T.~G.}\ \bibnamefont {Blackburn}}, \bibinfo {author} {\bibfnamefont
			{N.}~\bibnamefont {Bourgeois}}, \bibinfo {author} {\bibfnamefont
			{L.}~\bibnamefont {Calvin}}, \emph {et~al.},\ }\bibfield  {title} {\bibinfo
		{title} {Observation of quantum effects on radiation reaction in strong
			fields},\ }\href {https://doi.org/10.1038/s41467-025-67918-8} {\bibfield
		{journal} {\bibinfo  {journal} {Nature Commun.}\ }\textbf {\bibinfo {volume}
			{17}},\ \bibinfo {pages} {1157} (\bibinfo {year} {2026})}\BibitemShut
	{NoStop}%
	\bibitem [{\citenamefont {Lindl}(1995)}]{lindl1995development}%
	\BibitemOpen
	\bibfield  {author} {\bibinfo {author} {\bibfnamefont {J.}~\bibnamefont
			{Lindl}},\ }\bibfield  {title} {\bibinfo {title} {Development of the
			indirect-drive approach to inertial confinement fusion and the target physics
			basis for ignition and gain},\ }\href {https://doi.org/10.1063/1.871025}
	{\bibfield  {journal} {\bibinfo  {journal} {Phys. Plasmas}\ }\textbf
		{\bibinfo {volume} {2}},\ \bibinfo {pages} {3933} (\bibinfo {year}
		{1995})}\BibitemShut {NoStop}%
	\bibitem [{\citenamefont {Craxton}\ \emph {et~al.}(2015)\citenamefont
		{Craxton}, \citenamefont {Anderson}, \citenamefont {Boehly}, \citenamefont
		{Goncharov}, \citenamefont {Harding}, \citenamefont {Knauer}, \citenamefont
		{McCrory}, \citenamefont {McKenty}, \citenamefont {Meyerhofer}, \citenamefont
		{Myatt} \emph {et~al.}}]{craxton2015direct}%
	\BibitemOpen
	\bibfield  {author} {\bibinfo {author} {\bibfnamefont {R.~S.}\ \bibnamefont
			{Craxton}}, \bibinfo {author} {\bibfnamefont {K.~S.}\ \bibnamefont
			{Anderson}}, \bibinfo {author} {\bibfnamefont {T.~R.}\ \bibnamefont
			{Boehly}}, \bibinfo {author} {\bibfnamefont {V.~N.}\ \bibnamefont
			{Goncharov}}, \bibinfo {author} {\bibfnamefont {D.~R.}\ \bibnamefont
			{Harding}}, \bibinfo {author} {\bibfnamefont {J.~P.}\ \bibnamefont {Knauer}},
		\bibinfo {author} {\bibfnamefont {R.~L.}\ \bibnamefont {McCrory}}, \bibinfo
		{author} {\bibfnamefont {P.~W.}\ \bibnamefont {McKenty}}, \bibinfo {author}
		{\bibfnamefont {D.~D.}\ \bibnamefont {Meyerhofer}}, \bibinfo {author}
		{\bibfnamefont {J.~F.}\ \bibnamefont {Myatt}}, \emph {et~al.},\ }\bibfield
	{title} {\bibinfo {title} {Direct-drive inertial confinement fusion: A
			review},\ }\href {https://doi.org/10.1063/1.4934714} {\bibfield  {journal}
		{\bibinfo  {journal} {Phys. Plasmas}\ }\textbf {\bibinfo {volume} {22}},\
		\bibinfo {pages} {110501} (\bibinfo {year} {2015})}\BibitemShut {NoStop}%
	\bibitem [{\citenamefont {Born}(1909)}]{Born:1909rpp}%
	\BibitemOpen
	\bibfield  {author} {\bibinfo {author} {\bibfnamefont {M.}~\bibnamefont
			{Born}},\ }\bibfield  {title} {\bibinfo {title} {{The theory of the rigid
				electron in the kinematics of the relativity principle}},\ }\href
	{https://doi.org/10.1002/andp.19093351102} {\bibfield  {journal} {\bibinfo
			{journal} {Annalen Phys.}\ }\textbf {\bibinfo {volume} {30}},\ \bibinfo
		{pages} {1} (\bibinfo {year} {1909})}\BibitemShut {NoStop}%
	\bibitem [{\citenamefont {Rindler}(1960)}]{Rindler:1960zz}%
	\BibitemOpen
	\bibfield  {author} {\bibinfo {author} {\bibfnamefont {W.}~\bibnamefont
			{Rindler}},\ }\bibfield  {title} {\bibinfo {title} {{Hyperbolic Motion in
				Curved Space Time}},\ }\href {https://doi.org/10.1103/PhysRev.119.2082}
	{\bibfield  {journal} {\bibinfo  {journal} {Phys. Rev.}\ }\textbf {\bibinfo
			{volume} {119}},\ \bibinfo {pages} {2082} (\bibinfo {year}
		{1960})}\BibitemShut {NoStop}%
	\bibitem [{\citenamefont {Fulton}\ and\ \citenamefont
		{Rohrlich}(1960)}]{FultonRohrlich1960}%
	\BibitemOpen
	\bibfield  {author} {\bibinfo {author} {\bibfnamefont {T.}~\bibnamefont
			{Fulton}}\ and\ \bibinfo {author} {\bibfnamefont {F.}~\bibnamefont
			{Rohrlich}},\ }\bibfield  {title} {\bibinfo {title} {Classical radiation from
			a uniformly accelerated charge},\ }\href
	{https://doi.org/10.1016/0003-4916(60)90105-6} {\bibfield  {journal}
		{\bibinfo  {journal} {Ann. Phys.}\ }\textbf {\bibinfo {volume} {9}},\
		\bibinfo {pages} {499} (\bibinfo {year} {1960})}\BibitemShut {NoStop}%
	\bibitem [{\citenamefont {Boulware}(1980)}]{Boulware:1979qj}%
	\BibitemOpen
	\bibfield  {author} {\bibinfo {author} {\bibfnamefont {D.~G.}\ \bibnamefont
			{Boulware}},\ }\bibfield  {title} {\bibinfo {title} {{Radiation From a
				Uniformly Accelerated Charge}},\ }\href
	{https://doi.org/10.1016/0003-4916(80)90360-7} {\bibfield  {journal}
		{\bibinfo  {journal} {Ann. Phys.}\ }\textbf {\bibinfo {volume} {124}},\
		\bibinfo {pages} {169} (\bibinfo {year} {1980})}\BibitemShut {NoStop}%
	\bibitem [{\citenamefont {Rohrlich}(2007)}]{Rohrlich2007}%
	\BibitemOpen
	\bibfield  {author} {\bibinfo {author} {\bibfnamefont {F.}~\bibnamefont
			{Rohrlich}},\ }\href@noop {} {\emph {\bibinfo {title} {Classical Charged
				Particles}}},\ \bibinfo {edition} {3rd}\ ed.\ (\bibinfo  {publisher} {World
		Scientific},\ \bibinfo {address} {Singapore},\ \bibinfo {year}
	{2007})\BibitemShut {NoStop}%
	\bibitem [{\citenamefont {Eriksen}\ and\ \citenamefont
		{Gr{\o}n}(2000)}]{EriksenGron2000}%
	\BibitemOpen
	\bibfield  {author} {\bibinfo {author} {\bibfnamefont {E.}~\bibnamefont
			{Eriksen}}\ and\ \bibinfo {author} {\bibfnamefont {{\O}.}~\bibnamefont
			{Gr{\o}n}},\ }\bibfield  {title} {\bibinfo {title} {Electrodynamics of
			hyperbolically accelerated charges. {I}. {T}he electromagnetic field of a
			charged particle with hyperbolic motion},\ }\href
	{https://doi.org/10.1006/aphy.2000.6094} {\bibfield  {journal} {\bibinfo
			{journal} {Ann. Phys.}\ }\textbf {\bibinfo {volume} {286}},\ \bibinfo {pages}
		{320} (\bibinfo {year} {2000})}\BibitemShut {NoStop}%
	\bibitem [{\citenamefont {Teitelboim}(1970)}]{Teitelboim:1970mw}%
	\BibitemOpen
	\bibfield  {author} {\bibinfo {author} {\bibfnamefont {C.}~\bibnamefont
			{Teitelboim}},\ }\bibfield  {title} {\bibinfo {title} {{Splitting of the
				{M}axwell tensor - radiation reaction without advanced fields}},\ }\href
	{https://doi.org/10.1103/PhysRevD.2.1763} {\bibfield  {journal} {\bibinfo
			{journal} {Phys. Rev. D}\ }\textbf {\bibinfo {volume} {1}},\ \bibinfo {pages}
		{1572} (\bibinfo {year} {1970})}\BibitemShut {NoStop}%
	\bibitem [{\citenamefont {Taub}(1948)}]{Taub1948}%
	\BibitemOpen
	\bibfield  {author} {\bibinfo {author} {\bibfnamefont {A.~H.}\ \bibnamefont
			{Taub}},\ }\bibfield  {title} {\bibinfo {title} {Orbits of charged particles
			in constant fields},\ }\href {https://doi.org/doi.org/10.1103/PhysRev.73.786}
	{\bibfield  {journal} {\bibinfo  {journal} {Phys. Rev.}\ }\textbf {\bibinfo
			{volume} {73}},\ \bibinfo {pages} {786} (\bibinfo {year} {1948})}\BibitemShut
	{NoStop}%
	\bibitem [{\citenamefont {Dumais}(1985)}]{Dumais1985}%
	\BibitemOpen
	\bibfield  {author} {\bibinfo {author} {\bibfnamefont {J.-F.}\ \bibnamefont
			{Dumais}},\ }\bibfield  {title} {\bibinfo {title} {Eigenvectors of the
			electromagnetic field tensor and relativistic charged-particle motions},\
	}\href {https://doi.org/10.1119/1.14136} {\bibfield  {journal} {\bibinfo
			{journal} {Am. J. Phys.}\ }\textbf {\bibinfo {volume} {53}},\ \bibinfo
		{pages} {264} (\bibinfo {year} {1985})}\BibitemShut {NoStop}%
	\bibitem [{\citenamefont {Fradkin}(1978)}]{Fradkin:1978kx}%
	\BibitemOpen
	\bibfield  {author} {\bibinfo {author} {\bibfnamefont {D.~M.}\ \bibnamefont
			{Fradkin}},\ }\bibfield  {title} {\bibinfo {title} {{Covariant
				Electromagnetic Projection Operators and a Covariant Description of Charged
				Particle Guiding Center Motion}},\ }\href
	{https://doi.org/10.1088/0305-4470/11/6/010} {\bibfield  {journal} {\bibinfo
			{journal} {J. Phys. A}\ }\textbf {\bibinfo {volume} {11}},\ \bibinfo {pages}
		{1069} (\bibinfo {year} {1978})}\BibitemShut {NoStop}%
	\bibitem [{\citenamefont {Mu{\~n}oz}(1997)}]{Munoz1997}%
	\BibitemOpen
	\bibfield  {author} {\bibinfo {author} {\bibfnamefont {G.}~\bibnamefont
			{Mu{\~n}oz}},\ }\bibfield  {title} {\bibinfo {title} {Charged particle in a
			constant electromagnetic field: Covariant solution},\ }\href
	{https://doi.org/10.1119/1.18626} {\bibfield  {journal} {\bibinfo  {journal}
			{Am. J. Phys.}\ }\textbf {\bibinfo {volume} {65}},\ \bibinfo {pages} {429}
		(\bibinfo {year} {1997})}\BibitemShut {NoStop}%
	\bibitem [{\citenamefont {Chin}(2009)}]{Chin2009}%
	\BibitemOpen
	\bibfield  {author} {\bibinfo {author} {\bibfnamefont {S.~A.}\ \bibnamefont
			{Chin}},\ }\bibfield  {title} {\bibinfo {title} {Relativistic motion in a
			constant electromagnetic field},\ }\href {https://doi.org/10.1063/1.3064796}
	{\bibfield  {journal} {\bibinfo  {journal} {J. Math. Phys.}\ }\textbf
		{\bibinfo {volume} {50}},\ \bibinfo {pages} {012904} (\bibinfo {year}
		{2009})}\BibitemShut {NoStop}%
	\bibitem [{\citenamefont {Bulanov}\ \emph {et~al.}(2011)\citenamefont
		{Bulanov}, \citenamefont {Esirkepov}, \citenamefont {Kando}, \citenamefont
		{Koga},\ and\ \citenamefont {Bulanov}}]{Bulanov2011}%
	\BibitemOpen
	\bibfield  {author} {\bibinfo {author} {\bibfnamefont {S.~V.}\ \bibnamefont
			{Bulanov}}, \bibinfo {author} {\bibfnamefont {T.~Z.}\ \bibnamefont
			{Esirkepov}}, \bibinfo {author} {\bibfnamefont {M.}~\bibnamefont {Kando}},
		\bibinfo {author} {\bibfnamefont {J.~K.}\ \bibnamefont {Koga}},\ and\
		\bibinfo {author} {\bibfnamefont {S.~S.}\ \bibnamefont {Bulanov}},\
	}\bibfield  {title} {\bibinfo {title} {Lorentz-{A}braham-{D}irac versus
			{L}andau-{L}ifshitz radiation friction force in the ultrarelativistic
			electron interaction with electromagnetic wave (exact solutions)},\ }\href
	{https://doi.org/10.1103/PhysRevE.84.056605} {\bibfield  {journal} {\bibinfo
			{journal} {Phys. Rev. E}\ }\textbf {\bibinfo {volume} {84}},\ \bibinfo
		{pages} {056605} (\bibinfo {year} {2011})}\BibitemShut {NoStop}%
	\bibitem [{\citenamefont {Gonoskov}\ and\ \citenamefont
		{Marklund}(2018)}]{Gonoskov:2017lyz}%
	\BibitemOpen
	\bibfield  {author} {\bibinfo {author} {\bibfnamefont {A.}~\bibnamefont
			{Gonoskov}}\ and\ \bibinfo {author} {\bibfnamefont {M.}~\bibnamefont
			{Marklund}},\ }\bibfield  {title} {\bibinfo {title} {{Radiation-dominated
				particle and plasma dynamics}},\ }\href {https://doi.org/10.1063/1.5047799}
	{\bibfield  {journal} {\bibinfo  {journal} {Phys. Plasmas}\ }\textbf
		{\bibinfo {volume} {25}},\ \bibinfo {pages} {093109} (\bibinfo {year}
		{2018})}\BibitemShut {NoStop}%
	\bibitem [{\citenamefont {Samsonov}\ \emph {et~al.}(2018)\citenamefont
		{Samsonov}, \citenamefont {Nerush},\ and\ \citenamefont
		{Kostyukov}}]{Samsonov:2018skj}%
	\BibitemOpen
	\bibfield  {author} {\bibinfo {author} {\bibfnamefont {A.~S.}\ \bibnamefont
			{Samsonov}}, \bibinfo {author} {\bibfnamefont {E.~N.}\ \bibnamefont
			{Nerush}},\ and\ \bibinfo {author} {\bibfnamefont {I.~Y.}\ \bibnamefont
			{Kostyukov}},\ }\bibfield  {title} {\bibinfo {title} {{Asymptotic electron
				motion in the strongly-radiation-dominated regime}},\ }\href
	{https://doi.org/10.1103/PhysRevA.98.053858} {\bibfield  {journal} {\bibinfo
			{journal} {Phys. Rev. A}\ }\textbf {\bibinfo {volume} {98}},\ \bibinfo
		{pages} {053858} (\bibinfo {year} {2018})}\BibitemShut {NoStop}%
	\bibitem [{\citenamefont {Ekman}\ \emph {et~al.}(2021)\citenamefont {Ekman},
		\citenamefont {Heinzl},\ and\ \citenamefont {Ilderton}}]{Ekman:2021vwg}%
	\BibitemOpen
	\bibfield  {author} {\bibinfo {author} {\bibfnamefont {R.}~\bibnamefont
			{Ekman}}, \bibinfo {author} {\bibfnamefont {T.}~\bibnamefont {Heinzl}},\ and\
		\bibinfo {author} {\bibfnamefont {A.}~\bibnamefont {Ilderton}},\ }\bibfield
	{title} {\bibinfo {title} {{Exact solutions in radiation reaction and the
				radiation-free direction}},\ }\href
	{https://doi.org/10.1088/1367-2630/abfab2} {\bibfield  {journal} {\bibinfo
			{journal} {New J. Phys.}\ }\textbf {\bibinfo {volume} {23}},\ \bibinfo
		{pages} {055001} (\bibinfo {year} {2021})}\BibitemShut {NoStop}%
	\bibitem [{\citenamefont {Cai}\ \emph {et~al.}(2023{\natexlab{a}})\citenamefont
		{Cai}, \citenamefont {Gralla},\ and\ \citenamefont
		{Paschalidis}}]{Cai:2022mkw}%
	\BibitemOpen
	\bibfield  {author} {\bibinfo {author} {\bibfnamefont {Y.}~\bibnamefont
			{Cai}}, \bibinfo {author} {\bibfnamefont {S.~E.}\ \bibnamefont {Gralla}},\
		and\ \bibinfo {author} {\bibfnamefont {V.}~\bibnamefont {Paschalidis}},\
	}\bibfield  {title} {\bibinfo {title} {{Dynamics of ultrarelativistic charged
				particles with strong radiation reaction. I. Aristotelian equilibrium
				state}},\ }\href {https://doi.org/10.1103/PhysRevD.108.063018} {\bibfield
		{journal} {\bibinfo  {journal} {Phys. Rev. D}\ }\textbf {\bibinfo {volume}
			{108}},\ \bibinfo {pages} {063018} (\bibinfo {year}
		{2023}{\natexlab{a}})}\BibitemShut {NoStop}%
	\bibitem [{\citenamefont {Cai}\ \emph {et~al.}(2023{\natexlab{b}})\citenamefont
		{Cai}, \citenamefont {Gralla},\ and\ \citenamefont
		{Paschalidis}}]{Cai:2023ykg}%
	\BibitemOpen
	\bibfield  {author} {\bibinfo {author} {\bibfnamefont {Y.}~\bibnamefont
			{Cai}}, \bibinfo {author} {\bibfnamefont {S.~E.}\ \bibnamefont {Gralla}},\
		and\ \bibinfo {author} {\bibfnamefont {V.}~\bibnamefont {Paschalidis}},\
	}\bibfield  {title} {\bibinfo {title} {{Dynamics of ultrarelativistic charged
				particles with strong radiation reaction. II. Entry into Aristotelian
				equilibrium}},\ }\href {https://doi.org/10.1103/PhysRevD.108.063019}
	{\bibfield  {journal} {\bibinfo  {journal} {Phys. Rev. D}\ }\textbf {\bibinfo
			{volume} {108}},\ \bibinfo {pages} {063019} (\bibinfo {year}
		{2023}{\natexlab{b}})}\BibitemShut {NoStop}%
	\bibitem [{\citenamefont {Samsonov}\ \emph {et~al.}(2023)\citenamefont
		{Samsonov}, \citenamefont {Nerush},\ and\ \citenamefont
		{Kostyukov}}]{Samsonov2023}%
	\BibitemOpen
	\bibfield  {author} {\bibinfo {author} {\bibfnamefont {A.~S.}\ \bibnamefont
			{Samsonov}}, \bibinfo {author} {\bibfnamefont {E.~N.}\ \bibnamefont
			{Nerush}},\ and\ \bibinfo {author} {\bibfnamefont {I.~Y.}\ \bibnamefont
			{Kostyukov}},\ }\bibfield  {title} {\bibinfo {title} {High-order corrections
			to the radiation-free dynamics of an electron in the strongly
			radiation-dominated regime},\ }\href {https://doi.org/10.1063/5.0117504}
	{\bibfield  {journal} {\bibinfo  {journal} {Matter Radiat. Extremes}\
		}\textbf {\bibinfo {volume} {8}},\ \bibinfo {pages} {014402} (\bibinfo {year}
		{2023})}\BibitemShut {NoStop}%
	\bibitem [{\citenamefont {Yaremko}(2013)}]{Yaremko:2013eqe}%
	\BibitemOpen
	\bibfield  {author} {\bibinfo {author} {\bibfnamefont {Y.}~\bibnamefont
			{Yaremko}},\ }\bibfield  {title} {\bibinfo {title} {{Exact solution to the
				Landau-Lifshitz equation in a constant electromagnetic field}},\ }\href
	{https://doi.org/10.1063/1.4820131} {\bibfield  {journal} {\bibinfo
			{journal} {J. Math. Phys.}\ }\textbf {\bibinfo {volume} {54}},\ \bibinfo
		{pages} {092901} (\bibinfo {year} {2013})}\BibitemShut {NoStop}%
	\bibitem [{\citenamefont {Kazinski}\ and\ \citenamefont
		{Shipulya}(2011)}]{Kazinski:2010ce}%
	\BibitemOpen
	\bibfield  {author} {\bibinfo {author} {\bibfnamefont {P.~O.}\ \bibnamefont
			{Kazinski}}\ and\ \bibinfo {author} {\bibfnamefont {M.~A.}\ \bibnamefont
			{Shipulya}},\ }\bibfield  {title} {\bibinfo {title} {{Asymptotics of physical
				solutions to the Lorentz-Dirac equation for a planar motion in constant
				electromagnetic fields}},\ }\href
	{https://doi.org/10.1103/PhysRevE.83.066606} {\bibfield  {journal} {\bibinfo
			{journal} {Phys. Rev. E}\ }\textbf {\bibinfo {volume} {83}},\ \bibinfo
		{pages} {066606} (\bibinfo {year} {2011})}\BibitemShut {NoStop}%
	\bibitem [{\citenamefont {Kazinski}(2013)}]{Kazinski:2013vga}%
	\BibitemOpen
	\bibfield  {author} {\bibinfo {author} {\bibfnamefont {P.~O.}\ \bibnamefont
			{Kazinski}},\ }\bibfield  {title} {\bibinfo {title} {{Radiation of de-excited
				electrons at large times in a strong electromagnetic plane wave}},\ }\href
	{https://doi.org/10.1016/j.aop.2013.09.016} {\bibfield  {journal} {\bibinfo
			{journal} {Ann. Phys.}\ }\textbf {\bibinfo {volume} {339}},\ \bibinfo {pages}
		{430} (\bibinfo {year} {2013})}\BibitemShut {NoStop}%
	\bibitem [{\citenamefont {Rafelski}(2017)}]{Rafelski:2017hyt}%
	\BibitemOpen
	\bibfield  {author} {\bibinfo {author} {\bibfnamefont {J.}~\bibnamefont
			{Rafelski}},\ }\href {https://doi.org/10.1007/978-3-319-51231-0} {\emph
		{\bibinfo {title} {{Relativity Matters. From Einstein's EMC2 to Laser
					Particle Acceleration and Quark-Gluon Plasma}}}}\ (\bibinfo  {publisher}
	{Springer},\ \bibinfo {year} {2017})\BibitemShut {NoStop}%
	\bibitem [{\citenamefont {Schwinger}(1951)}]{Schwinger:1951nm}%
	\BibitemOpen
	\bibfield  {author} {\bibinfo {author} {\bibfnamefont {J.~S.}\ \bibnamefont
			{Schwinger}},\ }\bibfield  {title} {\bibinfo {title} {{On gauge invariance
				and vacuum polarization}},\ }\href {https://doi.org/10.1103/PhysRev.82.664}
	{\bibfield  {journal} {\bibinfo  {journal} {Phys. Rev.}\ }\textbf {\bibinfo
			{volume} {82}},\ \bibinfo {pages} {664} (\bibinfo {year} {1951})}\BibitemShut
	{NoStop}%
	\bibitem [{\citenamefont {Di~Piazza}(2008)}]{Piazza:2008yvh}%
	\BibitemOpen
	\bibfield  {author} {\bibinfo {author} {\bibfnamefont {A.}~\bibnamefont
			{Di~Piazza}},\ }\bibfield  {title} {\bibinfo {title} {{Exact Solution of the
				Landau-Lifshitz Equation in a Plane Wave}},\ }\href
	{https://doi.org/10.1007/s11005-008-0228-9} {\bibfield  {journal} {\bibinfo
			{journal} {Lett. Math. Phys.}\ }\textbf {\bibinfo {volume} {83}},\ \bibinfo
		{pages} {305} (\bibinfo {year} {2008})}\BibitemShut {NoStop}%
	\bibitem [{\citenamefont {Hadad}\ \emph {et~al.}(2010)\citenamefont {Hadad},
		\citenamefont {Labun}, \citenamefont {Rafelski}, \citenamefont {Elkina},
		\citenamefont {Klier},\ and\ \citenamefont {Ruhl}}]{Hadad:2010mt}%
	\BibitemOpen
	\bibfield  {author} {\bibinfo {author} {\bibfnamefont {Y.}~\bibnamefont
			{Hadad}}, \bibinfo {author} {\bibfnamefont {L.}~\bibnamefont {Labun}},
		\bibinfo {author} {\bibfnamefont {J.}~\bibnamefont {Rafelski}}, \bibinfo
		{author} {\bibfnamefont {N.}~\bibnamefont {Elkina}}, \bibinfo {author}
		{\bibfnamefont {C.}~\bibnamefont {Klier}},\ and\ \bibinfo {author}
		{\bibfnamefont {H.}~\bibnamefont {Ruhl}},\ }\bibfield  {title} {\bibinfo
		{title} {{Effects of Radiation-Reaction in Relativistic Laser
				Acceleration}},\ }\href {https://doi.org/10.1103/PhysRevD.82.096012}
	{\bibfield  {journal} {\bibinfo  {journal} {Phys. Rev. D}\ }\textbf {\bibinfo
			{volume} {82}},\ \bibinfo {pages} {096012} (\bibinfo {year}
		{2010})}\BibitemShut {NoStop}%
	\bibitem [{\citenamefont {Ritus}(1985)}]{Ritus:1985vta}%
	\BibitemOpen
	\bibfield  {author} {\bibinfo {author} {\bibfnamefont {V.~I.}\ \bibnamefont
			{Ritus}},\ }\bibfield  {title} {\bibinfo {title} {{Quantum effects of the
				interaction of elementary particles with an intense electromagnetic field}},\
	}\href {https://doi.org/10.1007/BF01120220} {\bibfield  {journal} {\bibinfo
			{journal} {J. Russ. Laser Res.}\ }\textbf {\bibinfo {volume} {6}},\ \bibinfo
		{pages} {497} (\bibinfo {year} {1985})}\BibitemShut {NoStop}%
	\bibitem [{\citenamefont {Caldirola}(1956)}]{caldirola1956new}%
	\BibitemOpen
	\bibfield  {author} {\bibinfo {author} {\bibfnamefont {P.}~\bibnamefont
			{Caldirola}},\ }\bibfield  {title} {\bibinfo {title} {A new model of
			classical electron},\ }\href {https://doi.org/10.1007/BF02743686} {\bibfield
		{journal} {\bibinfo  {journal} {Il Nuovo Cimento}\ }\textbf {\bibinfo
			{volume} {3}},\ \bibinfo {pages} {297} (\bibinfo {year} {1956})}\BibitemShut
	{NoStop}%
	\bibitem [{\citenamefont {Formanek}\ \emph {et~al.}(2020)\citenamefont
		{Formanek}, \citenamefont {Steinmetz},\ and\ \citenamefont
		{Rafelski}}]{Formanek:2020zwc}%
	\BibitemOpen
	\bibfield  {author} {\bibinfo {author} {\bibfnamefont {M.}~\bibnamefont
			{Formanek}}, \bibinfo {author} {\bibfnamefont {A.}~\bibnamefont
			{Steinmetz}},\ and\ \bibinfo {author} {\bibfnamefont {J.}~\bibnamefont
			{Rafelski}},\ }\bibfield  {title} {\bibinfo {title} {{Radiation reaction
				friction: Resistive material medium}},\ }\href
	{https://doi.org/10.1103/PhysRevD.102.056015} {\bibfield  {journal} {\bibinfo
			{journal} {Phys. Rev. D}\ }\textbf {\bibinfo {volume} {102}},\ \bibinfo
		{pages} {056015} (\bibinfo {year} {2020})}\BibitemShut {NoStop}%
\end{thebibliography}
%

\end{document}